\documentclass[epj]{webofc}
\usepackage[varg]{txfonts}   
\usepackage{graphicx}
\usepackage{amsmath}
\usepackage{bm}
\woctitle{POETIC6}
\begin{document}
\title{Tagged spectator deep-inelastic scattering off the deuteron as a tool to 
study neutron structure}
%
%

\author{Wim Cosyn\inst{1}\fnsep\thanks{\email{wim.cosyn@ugent.be}} \and
        Misak Sargsian\inst{2}\fnsep\thanks{\email{sargsian@fiu.edu}} 
}

\institute{Department of Physics and Astronomy, Proeftuinstraat 86, Ghent 
University, 9000 Ghent, Belgium
\and
           Department of Physics, Florida International University, Miami FL 
33199, U.S.A.
          }

\abstract{ We give an overview of a model to describe deep-inelastic scattering 
(DIS) off the 
deuteron with a spectator proton, based on the virtual nucleon 
approximation 
(VNA).  The model accounts for the final-state 
interactions (FSI) of the DIS debris with the spectator proton.  Values of the 
rescattering cross section are obtained by fits to high-momentum spectator 
data.  By using the so-called ``pole extrapolation'' method, free neutron 
structure functions can be obtained by extrapolating low-momentum spectator 
proton data to the on-shell neutron pole.  We apply this method to the BONuS 
data set and find a surprising Bjorken $x$ dependence, indicating a possible 
rise of the neutron to proton structure function ratio at high $x$.}
\maketitle
\section{Introduction}
Tagged proton spectator DIS off the deuteron ($e+d \rightarrow e'+X+p_s$) can 
be used 
in several ways to study the role of QCD dynamics at nucleonic length scales.  
At low spectator momenta, one has access to the structure functions of an 
almost on-shell neutron.  Together with proton structure functions, flavor 
separation can be performed and the $d$ and $u$ quark densities can be 
determined.  At higher proton spectator momenta, the deuteron is in a 
high-density configuration and medium modifications of the structure functions 
can be quantified.  In kinematics where FSI mechanisms are large, one 
can use the reaction to constrain models for the $X$--proton rescattering.  
Recently, two experiments at Jefferson Lab have measured the reaction in 
different kinematics: Deeps at higher spectator momenta 
of a few hundred MeV~\cite{Klimenko:2005zz} 
and, BONuS at lower ones around 100 MeV~\cite{Baillie:2011za}.  In the near 
future, the 
BONuS12 experiment will run \cite{Bonus12} and the reaction is considered as a 
method to extract neutron structure at a future electron-ion collider 
\cite{Cosyn:2014zfa}.  In order to provide meaningful interpretations of the 
measured data, theoretical models that quantify the importance of the FSI in 
the reaction are needed 
\cite{Simula:1996xk,Melnitchouk:1996vp,Sargsian:2005rm,CiofidegliAtti:1999kp,
CiofidegliAtti:2002as,Palli:2009it}. The major 
difficulty in doing this is that one lacks detailed information about the 
composition and space-time evolution of the hadronic system produced in the 
deep-inelastic scattering and how this changes as a function of Bjorken $x$ and 
$Q^2$. Here, we present work in a model that has been used to account for FSI 
mechanisms at moderate and high Bjorken $x$ in $e+d \rightarrow e'+X+p_s$ 
\cite{Cosyn:2010ux,Cosyn:2011jc,Cosyn:2015mha} as well as inclusive deuteron DIS 
\cite{Cosyn:2013uoa,Cosyn:2014sqa}.

\section{Formalism}
The model introduced in Ref.~\cite{Cosyn:2010ux} accounted
for FSI effects in DIS from the deuteron based on general properties of 
high-energy diffractive scattering. The underlying assumption was that due to 
the restricted
phase space (finite values of $W$ and $Q^2$), the minimal Fock state
component of the wave function can be used to describe DIS from the
bound nucleon.  In this case the scattered state consists of three
outgoing valence quarks.  The model uses the virtual nucleon approximation 
(VNA) \cite{Melnitchouk:1996vp,Sargsian:2005rm,Sargsian:2001gu} and only 
considers the proton-neutron component of the deuteron wave function.  The 
effects of FSI are encoded in an effective rescattering amplitude between the 
produced hadronic mass $X$ and the spectator proton.  Given the GeV range of 
the energies in the experiments, the rescatterings will occur over small 
angles and be highly diffractive.  This allows us to model the rescattering 
amplitude using the generalized eikonal approximation (GEA) 
\cite{Frankfurt:1996xx}.  In the GEA, the eikonal rescattering amplitude 
$f_{X^\prime N, X N}$ is parametrized using three parameters (total 
rescattering cross section $\sigma_{XN}$, slope factor $B$, and real part 
$\epsilon$), that have a dependence on the $Q^2$ and Bjorken $x$ (or 
equivalently invariant mass $W$ of hadronic products $X$):
\begin{equation} \label{eq:fsiamp}
 f_{X^\prime N, X N} = \sigma(Q^2,x)\left[i + \epsilon(Q^2,
x)\right]e^{\frac{B(Q^2,x)}{2} t}.
\end{equation}
The cross sections including FSI are obtained in a factorized approach, whereby 
the DIS interaction of the virtual photon with the off-shell nucleon (encoded 
in the structure functions $F_{iN}$) is taken out of the integration over the 
momentum of the on-shell intermediate spectator nucleon.  Finally, the four 
structure functions of the tagged spectator DIS process can be 
written in the following form
\begin{equation}
 F_i^{SI}=\left(\alpha(i)F_{1N}+\beta(i)F_{2N} 
\right)S^D(p_s)(2\pi)^32E_s,\qquad 
\text{with} \, i=\text{L,T,LT,TT}.
\end{equation}
Here, $\alpha(i),\beta(i)$ are kinematical factors (see 
Ref.~\cite{Cosyn:2010ux} for detailed expressions), and $S^D(p_s)$ is the 
distorted deuteron momentum distribution that contains an impulse approximation 
(IA) contribution and an FSI contribution including the rescattering amplitude 
of Eq.~(\ref{eq:fsiamp})~\cite{Cosyn:2010ux}. 

Even for very low spectator momenta, the struck neutron is still off-shell in 
the deuteron.  One can still perform an analytical continuation, however, of 
the 
amplitude into the unphysical kinematical region and extrapolate it to 
the on-shell point $p_i^2=(p_D-p_s)^2=m_n^2$.  The IA part of the amplitude has 
a pole at this value of $p_i^2$, while the FSI part does not due to the 
additional integration over the loop momentum.  This is the so-called ``no-loop 
theorem''~\cite{Sargsian:2005rm}.  This means carrying out the pole 
extrapolation 
automatically gets rid of nuclear effects and allows one to extract the 
on-shell neutron structure function.  This procedure is similar to the ones 
first proposed by Chew and Lowin hadron-proton scattering~\cite{Chew:1958wd} .  
Due to the small binding energy of the deuteron, the actual extrapolation 
length to the on-shell pole is very small, and there are also no issues 
regarding a zero crossing for $p_i^2$. The pole extrapolation procedure for the 
extraction of $F_{2n}$ consists 
of multiplying 
the measured deuteron structure function, 
$F^{SI,\text{EXP}}_{2D}$ by the factor 
$I(\alpha_s,\bm p_{s\perp},t)$~\cite{Sargsian:2005rm}, which 
cancels the singularity of the IA amplitude and is normalized such that
\begin{equation}\label{eq:extrap}
F^{\text{extr}}_{2n}(Q^2,x,p_i^2) = I(\alpha_s, \bm p_{s\perp},p_i^2)\cdot 
F^{SI,\text{EXP}}_{2D}(Q^2,x,\alpha_s, \bm p_{s\perp})
\end{equation}
approaches the free $F_{2n}(Q^2,x)$ in the  
$p_i^2\rightarrow m_n^2$  limit with FSI effects being diminished.

\section{Results}
\begin{figure}
\sidecaption
\centering
\includegraphics[width=0.6\textwidth]{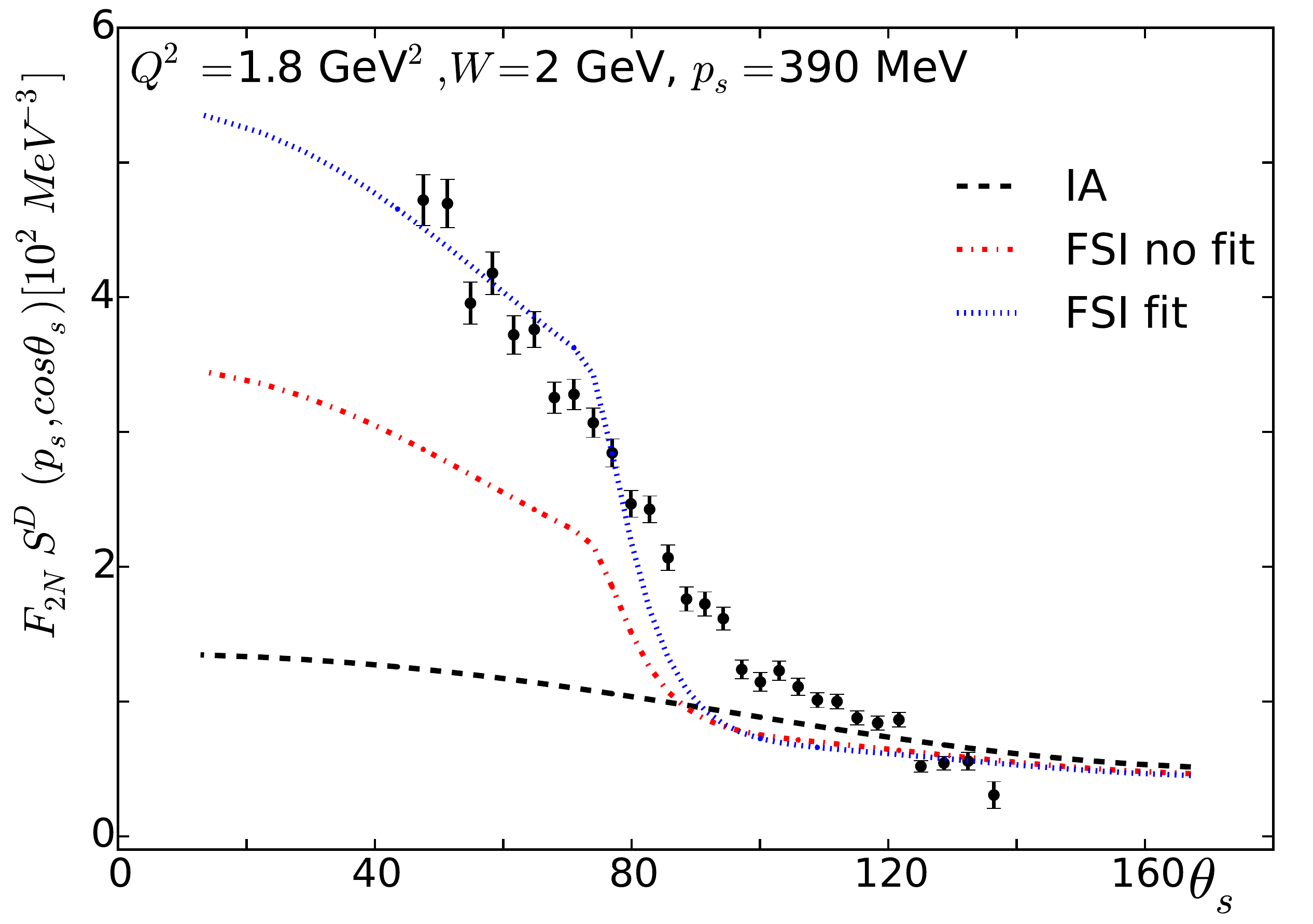}
\caption{Comparison between deeps data \cite{Klimenko:2005zz}, the plane-wave 
IA calculation (dashed black curve), an unfitted FSI calculation using standard 
$NN$ scattering values of $\sigma=40$ mb, $B=8$ GeV$^{-2}$, 
$\epsilon=-0.5$ (red dash-dotted), and a FSI calculation where $\sigma=55.8$ was
mb was fitted (blue dotted).}
\label{fig:nofit}
\end{figure}

Figure~\ref{fig:nofit} shows a comparison between one data set from the 
high spectator momentum deeps experiment~\cite{Klimenko:2005zz} with our model 
calculations.  The most striking feature of the data is the large excess at 
forward spectator angles.  This is markedly different from the quasi-elastic 
breakup case where the cross section peaks in the perpendicular 
region~\cite{Sargsian:2009hf}.  It is clear the IA calculation has an almost 
flat dependence on the spectator angle and can in no way reproduce the full 
range of the data.  An FSI calculation with values of the rescattering 
parameters inspired by the known $NN$ ones is able to reproduce the angular 
dependence better but still underestimates the data.  Intuitively, this is no 
surprise as for the value of $W=2$~GeV in Fig.~\ref{fig:nofit}, we are well 
above the nucleon mass.  Therefore, we have fitted a value of $\sigma$ for each 
$\{W,Q^2\}$ setting of the data.  The calculation using this fitted parameter is 
also shown for the kinematics of 
Fig.~\ref{fig:nofit}.  For a comparison with the full deeps data set, we refer 
to Ref.~\cite{Cosyn:2010ux}.

\begin{figure}
\sidecaption
\centering
\includegraphics[width=0.6\textwidth]{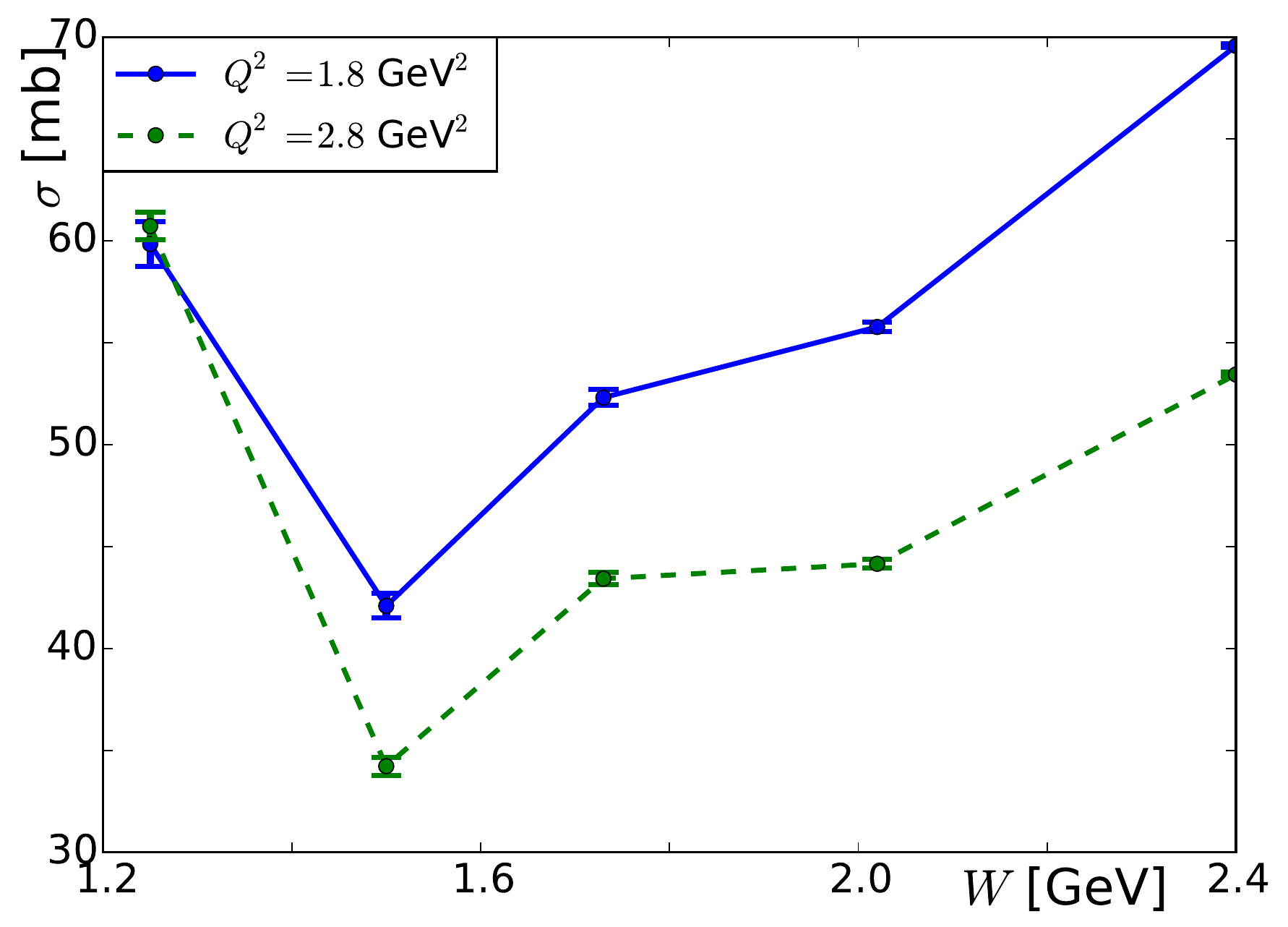}
\caption{Values of the fitted $\sigma$ rescattering parameter in 
Eq.~(\ref{eq:fsiamp}) for $Q^2=1.8~\text{GeV}^2$ (dashed green curve) and 
$Q^2=2.8~\text{GeV}^2$ (full blue curve), obtained by fitting the high momentum 
spectator deeps data.}
\label{fig:sigmafit}
\end{figure}

In Fig.~\ref{fig:sigmafit}, we show the $W$ and $Q^2$ dependence of the 
fitted values of $\sigma$ from our analysis of the deeps data.  The parameter 
$\sigma$ has a value of about 60 mb at the $\Delta$-resonance, and then clearly 
increases with increasing invariant mass $W$ of the hadronic products.  More 
interesting is the dependence on $Q^2$, where there is a significant decrease 
in the $\sigma$ parameter with increasing $Q^2$.  This could be a sign of a 
color transparency signal in this reaction, where the hadronic products are 
produced in a colorless point-like configuration at higher momentum transfers, 
which experiences reduced strong interactions with the remaining hadrons.  
More measurements at higher $Q^2$ values are needed, however, to further 
investigate this behavior.  

\begin{figure}
\sidecaption
\centering
\includegraphics[width=0.6\textwidth]{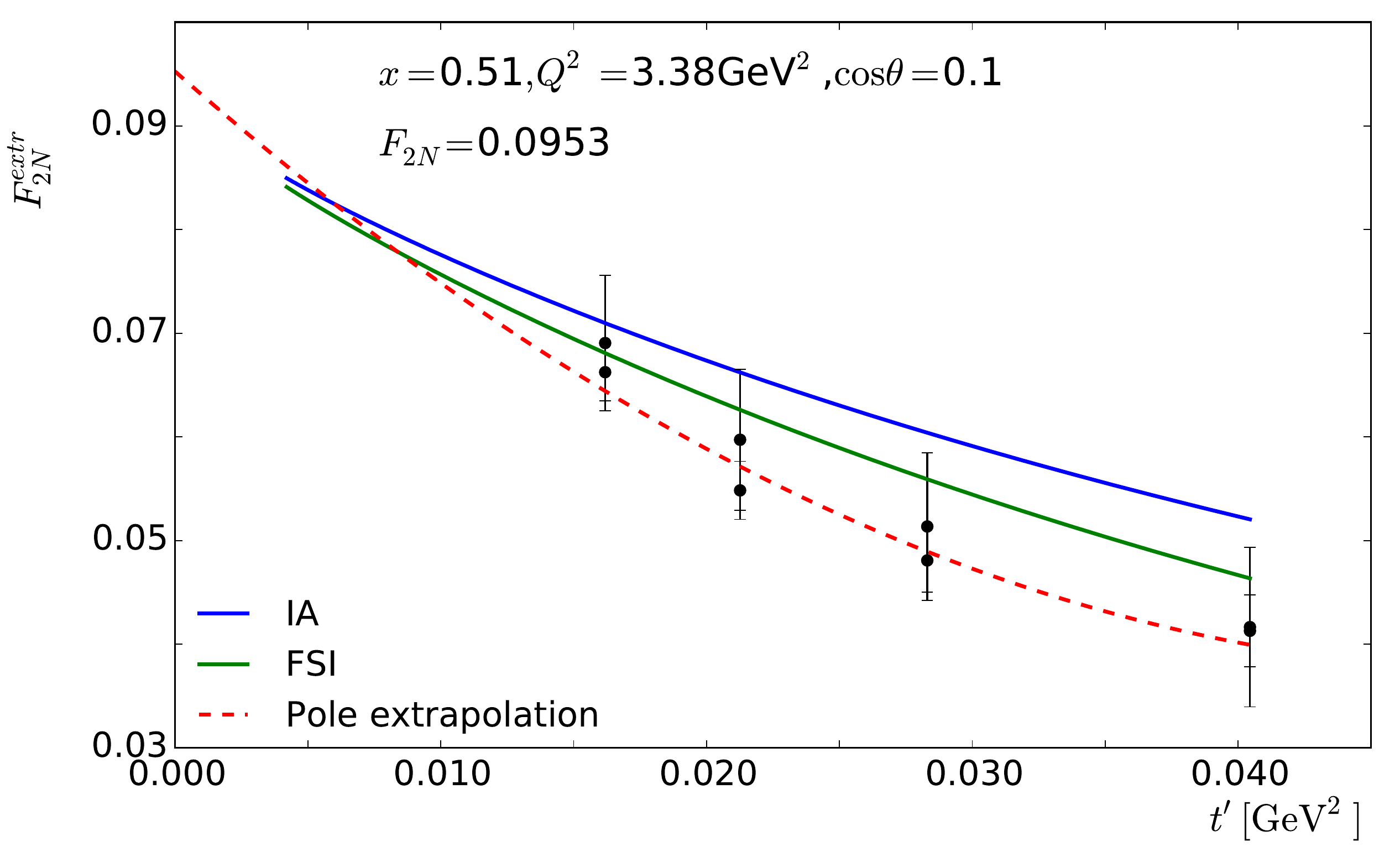}
\caption{Example of the pole extrapolation method using the renormalized BONuS 
data (black circles) with the quadratic pole extrapolation curve (red dashed 
curve) as a function of $t'^2=p_i^2-m_n^2$.  The IA (full blue curve) and FSI 
(full green curve) calculations are shown for comparison.}
\label{fig:extraponeangle}
\end{figure}

In Ref.~\cite{Cosyn:2011jc}, the pole extrapolation method was applied to the 
high momentum spectator deeps data.  Given the large extrapolation length, no 
robust results were obtained for $F_{2n}$.  With the more recent low spectator 
momentum BONuS data, which measured down to proton momenta of 70 MeV, the 
extrapolation length is very short and pole extrapolation became applicable.  
One problem with the BONuS data is that the measurements for different spectator 
momenta happened at different and not well known efficiencies.  In the original 
analysis, the data was therefore normalized to an IA model 
\cite{Tkachenko:2014byy} in the backward spectator region.  In 
Ref.~\cite{Cosyn:2015mha}, we renormalized the BONuS data using our FSI model 
at high $W,Q^2$ (where the neutron structure function is quite well known), 
using the rescattering parameter values extracted in our previous fit to the 
the 
deeps data.  Taking these renormalized data, we implemented the pole 
extrapolation method for the BONuS data.  Figure~\ref{fig:extraponeangle} shows 
an example for one kinematical setting.  The two data points at each $t'$ value 
correspond to data taken at two initial beam energy values and one can clearly 
observe they are consistent with each other.  Plotting our model calculations 
with the data, it is also clear that through implementing 
Eq.~(\ref{eq:extrap}), FSI effects indeed become smaller as one 
moves closer to the on-shell pole.  As the 
extrapolation curve shows in Fig.~\ref{fig:extraponeangle}, the extrapolation 
distance to the on-shell point is reasonable.

\begin{figure}
\sidecaption
\centering
\includegraphics[width=0.6\textwidth]{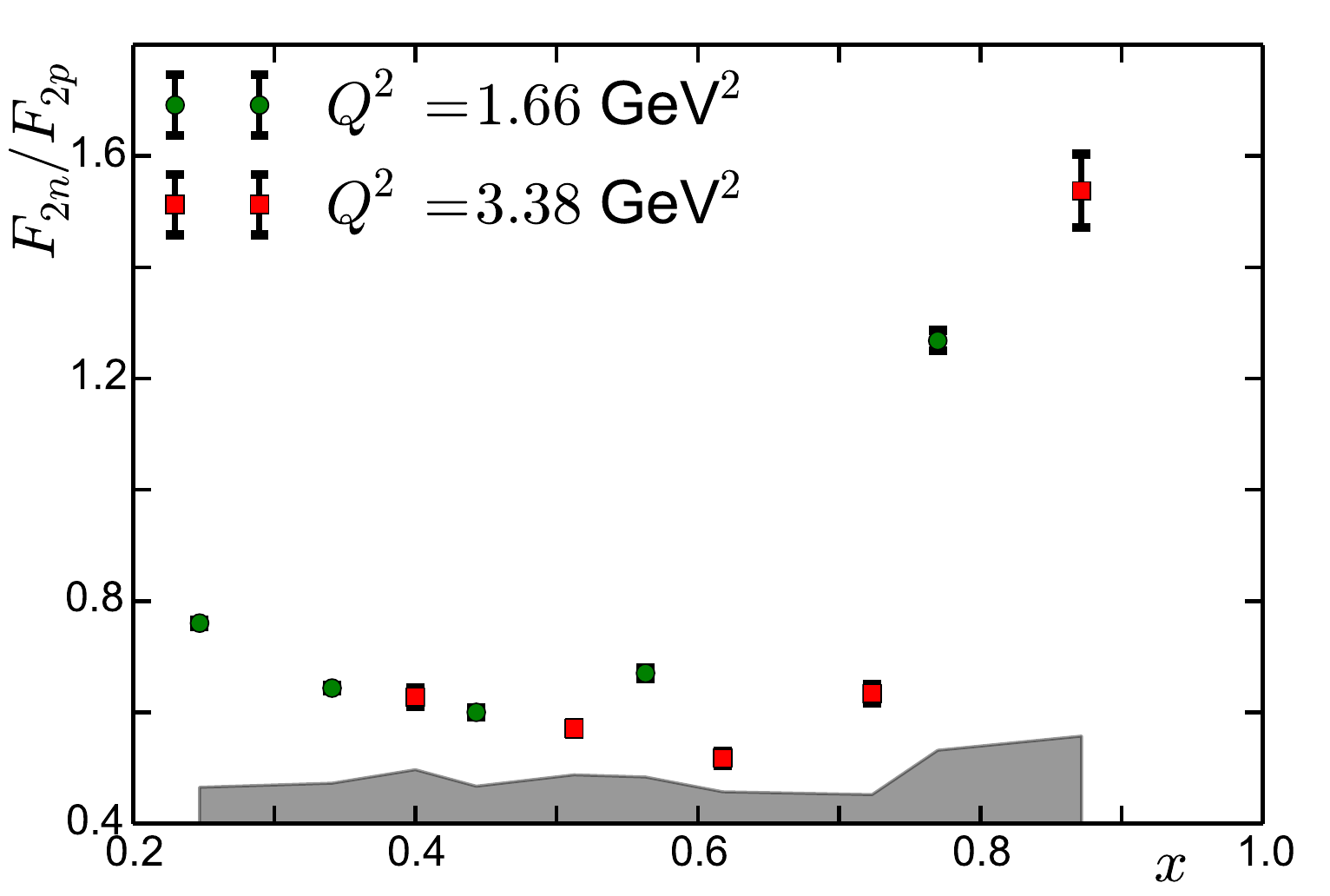}
\caption{$F_{2n}$ to $F_{2p}$ ratio obtained using the
pole extrapolation applied to the renormalized BONuS data.  The grey band 
denotes estimated systematic errors in the procedure.}
\label{fig:f2nf2p}
\end{figure}

For each kinematical setting of the BONuS data, pole extrapolation was carried 
out as in Fig.~\ref{fig:extraponeangle}, and a weighted average was taken over 
all measured spectator angle bins.  The final result is shown in 
Fig.~\ref{fig:f2nf2p} as the 
ratio of $F_{2n}/F_{2p}$ where the parametrization of 
Ref.~\cite{Christy:2007ve} was used for the $F_{2p}$ values.   We repeat 
that the values obtained in this manner are free of Fermi motion and nuclear 
medium modification effects. The systematic 
errors were estimated by taking all uncertainties in the data and our 
normalization procedure (deuteron wave function, structure function 
parametrization, renormalization fit) into account in a Monte Carlo 
simulation~\cite{Cosyn:2015mha}.  The most striking feature in 
Fig.~\ref{fig:f2nf2p} is the rise of the ratio at large Bjorken $x$.  The 
results at lower Bjorken $x<0.5$ are in agreement with existing estimates.  It 
is worth noting that the upward trend of the ratio at higher $x$ is not due to 
our renormalization of the data.  Indeed, the values of the ratio are higher 
without the renormalization applied.  As the $W$ ($\approx 1.18~\text{GeV}$) 
values corresponding to 
the highest $x$ values  in Fig.~\ref{fig:f2nf2p} are in 
the sub-DIS regime, one cannot directly relate the rise of the ratio to the 
underlying properties of the $u$ and $d$ quark pdf's.  Such a connection can 
only be made using duality arguments.  It is worth noting, however, that the 
duality analysis of the BONuS data~\cite{Niculescu:2015wka}, concluded that  the 
$\Delta$-resonance contributes to the duality, 
within 20-30\% accuracy.  One possible explanation of the rise could be the 
presence of isosinglet 
$qq$ short-range correlations~(SRCs) in the nucleon at $x\rightarrow 1$.  Such a 
correlation will 
result in the same momentum sharing effect, 
which  is observed  recently  in asymmetric nuclei in the $NN$ SRC
region~\cite{Sargsian:2012sm}.

\section{Conclusion}
We have presented a selection of results obtained in a model used to describe 
the tagged spectator DIS process off the deuteron including the effect of FSI.  
We obtained a good description of data at higher spectator momenta and used 
these data to fit the $Q^2$- and $W$-dependence of the rescattering parameter 
$\sigma$.  For data at lower spectator momenta, we applied the pole 
extrapolation method, which allows to extract the free neutron structure 
function 
in a way that eliminates nuclear effects.  We observe an intriguing rise of the 
$F_{2n}/F_{2p}$ ratio at high Bjorken $x$.

\begin{acknowledgement}
The computational resources (Stevin Supercomputer Infrastructure) and
services used in this work were provided by Ghent University, the
Hercules Foundation and the Flemish Government – department EWI.  This
work is supported by the Research Foundation Flanders as well as by the  
U.S. Department of Energy Grant 
under Contract DE-FG02-01ER41172.
\end{acknowledgement}

%
\bibliography{../bibtexall.bib}

\begin{thebibliography}{23}

\bibitem{Klimenko:2005zz}
A.V. Klimenko et~al. (CLAS), Phys. Rev. \textbf{C73}, 035212 (2006),
  \texttt{nucl-ex/0510032}

\bibitem{Baillie:2011za}
N.~Baillie, S.~Tkachenko, J.~Zhang, P.~Bosted, S.~B\"ultmann, M.E. Christy,
  H.~Fenker, K.A. Griffioen, C.E. Keppel, S.E. Kuhn et~al. (CLAS
  Collaboration), Phys.Rev.Lett. \textbf{108}, 199902 (2012),
  \texttt{1110.2770}

\bibitem{Bonus12}
S.~Bueltmann (2006), jLAB-PR12-06-113

\bibitem{Cosyn:2014zfa}
W.~Cosyn, V.~Guzey, D.W. Higinbotham, C.~Hyde, S.~Kuhn, P.~Nadel-Turonski,
  K.~Park, M.~Sargsian, M.~Strikman, C.~Weiss, J. Phys. Conf. Ser.
  \textbf{543}, 012007 (2014), \texttt{1409.5768}

\bibitem{Simula:1996xk}
S.~Simula, Phys. Lett. \textbf{B387}, 245 (1996), \texttt{nucl-th/9605024}

\bibitem{Melnitchouk:1996vp}
W.~Melnitchouk, M.~Sargsian, M.I. Strikman, Z. Phys. \textbf{A359}, 99 (1997),
  \texttt{nucl-th/9609048}

\bibitem{Sargsian:2005rm}
M.~Sargsian, M.~Strikman, Phys. Lett. \textbf{B639}, 223 (2006),
  \texttt{hep-ph/0511054}

\bibitem{CiofidegliAtti:1999kp}
C.~Ciofi~degli Atti, L.P. Kaptari, S.~Scopetta, Eur. Phys. J. \textbf{A5}, 191
  (1999), \texttt{hep-ph/9904486}

\bibitem{CiofidegliAtti:2002as}
C.~Ciofi~degli Atti, B.Z. Kopeliovich, Eur. Phys. J. \textbf{A17}, 133 (2003),
  \texttt{nucl-th/0207001}

\bibitem{Palli:2009it}
V.~Palli, C.~Ciofi~degli Atti, L.P. Kaptari, C.B. Mezzetti, M.~Alvioli, Phys.
  Rev. \textbf{C80}, 054610 (2009), \texttt{0911.1377}

\bibitem{Cosyn:2010ux}
W.~Cosyn, M.~Sargsian, Phys. Rev. \textbf{C84}, 014601 (2011),
  \texttt{1012.0293}

\bibitem{Cosyn:2011jc}
W.~Cosyn, M.~Sargsian, AIP Conf.Proc. \textbf{1369}, 121 (2011),
  \texttt{1101.1258}

\bibitem{Cosyn:2015mha}
W.~Cosyn, M.M. Sargsian (2015), \texttt{1506.01067}

\bibitem{Cosyn:2013uoa}
W.~Cosyn, W.~Melnitchouk, M.~Sargsian, Phys.Rev. \textbf{C89}, 014612 (2014),
  \texttt{1311.3550}

\bibitem{Cosyn:2014sqa}
W.~Cosyn, M.~Sargsian, J. Phys. Conf. Ser. \textbf{543}, 012006 (2014),
  \texttt{1407.1653}

\bibitem{Sargsian:2001gu}
M.M. Sargsian, S.~Simula, M.I. Strikman, Phys. Rev. \textbf{C66}, 024001
  (2002), \texttt{nucl-th/0105052}

\bibitem{Frankfurt:1996xx}
L.L. Frankfurt, M.M. Sargsian, M.I. Strikman, Phys. Rev. \textbf{C56}, 1124
  (1997), \texttt{nucl-th/9603018}

\bibitem{Chew:1958wd}
G.F. Chew, F.E. Low, Phys. Rev. \textbf{113}, 1640 (1959)

\bibitem{Sargsian:2009hf}
M.M. Sargsian, Phys. Rev. \textbf{C82}, 014612 (2010), \texttt{0910.2016}

\bibitem{Tkachenko:2014byy}
S.~Tkachenko et~al. (CLAS), Phys. Rev. \textbf{C89}, 045206 (2014), [Addendum:
  Phys. Rev.C90,059901(2014)], \texttt{1402.2477}

\bibitem{Christy:2007ve}
M.~Christy, P.E. Bosted, Phys.Rev. \textbf{C81}, 055213 (2010),
  \texttt{0712.3731}

\bibitem{Niculescu:2015wka}
I.~Niculescu et~al., Phys. Rev. \textbf{C91}, 055206 (2015),
  \texttt{1501.02203}

\bibitem{Sargsian:2012sm}
M.M. Sargsian, Phys. Rev. \textbf{C89}, 034305 (2014), \texttt{1210.3280}

\end{thebibliography}

\end{document}